# Machine Learning and Bioinformatics for Diagnosis Analysis of Obesity Spectrum Disorders

Amin Gasmi

20 Mars 2020

**Abstract:**

Globally, the number of obese patients has doubled due to sedentary lifestyles and improper dieting. The tremendous increase altered human genetics, and health. According to the world health organization, Life expectancy dropped from 80 to 75 years, as obese people struggle with different chronic diseases. This report will address the problems of obesity in children and adults using ML datasets to feature, predict, and analyze the causes of obesity. By engaging neural ML networks, we will explore neural control using diffusion tensor imaging to consider body fats, BMI, waist & hip ratio circumference of obese patients. To predict the present and future causes of obesity with ML, we will discuss ML techniques like decision trees, SVM, RF, GBM, LASSO, BN, and ANN and use datasets implement the stated algorithms. Different theoretical literature from experts ML & Bioinformatics experiments will be outlined in this report while making recommendations on how to advance ML for predicting obesity and other chronic diseases.

**Introduction:**

Over the last five decades, clinicians have reported a 55 % increase in obese children & adults with a 39 % rate for developed countries and 16% from developing countries struggling with low-income and poor economic growth. In 2015, 32million children around the world were considered obese; specifically, children aged 5-15years. Records indicate potential struggles with metabolic syndromes that can cause diabetes, cardiovascular risk, and possible death. An adequate justification made on preventative practice was to reduce excess adipose tissues in the body. Hence, the fatty multifactorial etiological tissue causing obesity were presumed to be genetic, or due to lack of exercise and sedentary lifestyles. (Rankin et al.2020) [59]

The body mass index can impact and change human height & weight with a rebounding increasing & decrease in adipose tissue leading to tagged obese deciles. Obesity has no universal percentiles because an adult weighing 30kg/lm2 may be considered "obese". Obesity comes with either metabolic or hormonal changes and sometimes with pre-gestational factors triggering childhood obesity. Research further subjects childhood & adolescence obesity to parents' gestational weight gain, mother's lifestyle, and psychological factors during pregnancy. (González-Muniesa et al.2017)[60].

Diverse predictive models are engaged to detect risks, preventive measures, cost-effective methods of intervening in weight reduction projects. Based on statistical ranking projecting weight gain rates around the globe, simulation models are used to define sub-populated sets, age, and diet of obese individuals. Machine learning as part of the simulation techniques can be used to compare a high-dimensional & complex relationship between domain variables. Some ML tools are complex, nonlinear, automated, and predictive; especially deep learning ml with complex image, time-series, and text data model capacity. DT (Decision Tree) can be used to predict BMI in children & adults through collective patterns, statistical models, and outcomes that exhibit weight increase or decrease after therapy, physical fitness, and social commitment.

**Machine Learning tools**

The fundamental use of ML tools to predict data commence the late 90s when statistics & AL explored in developed countries. Rosenblat's perceptrons were seen as 1-layer ANN for performing baseline binary classification before NB, DT and KNN appeared to implement multilayer neural network backpropagated tasks. However, SVM, RF, and GBM emerged to foster CPU miniaturization due to its computational ability to manipulate new statistical models ML comprises of convolutional and recurrent layers that permit the training of huge datasets, data randomization, and complex training of million data parameters.

Whether the data parameters comprises of image, sound, text, or music; ML patterns with its feature map can handle all with its multilayer fully connected perceptron. Though, ML was perceived to be limited due to its traditional statistical models ,but aids medical & engineering approaches to data handling. To predict obesity with ML supervised/unsupervised models are endorsed to predict large variables or labeled data; whilst training with supervised models to identify age, sex, parent's BMI, and diet of obese individuals. To predict whether a child or an adult is likely to be obese in the future, an unsupervised model attempts to discover without labeled data, noise, and visualization; the probability of people likely to face obesity in the future. These grouped inputs are categorized with dimensioned reduction and clustering techniques.

Supervised ML uses two models to categorize datasets (classification & regression models) and precisely adopt numeric labeling, binary classification of "+" and "-" to reveal outputs of probability (p) of the given class of datasets. For instance, if "+" has an alternative comparable class of "-", the outcome equals 1-p, the threshold "t" will probably be equated as P>t, but if the same probability is assigned to its alternative category, the equation will change to P<t for the "-" category.

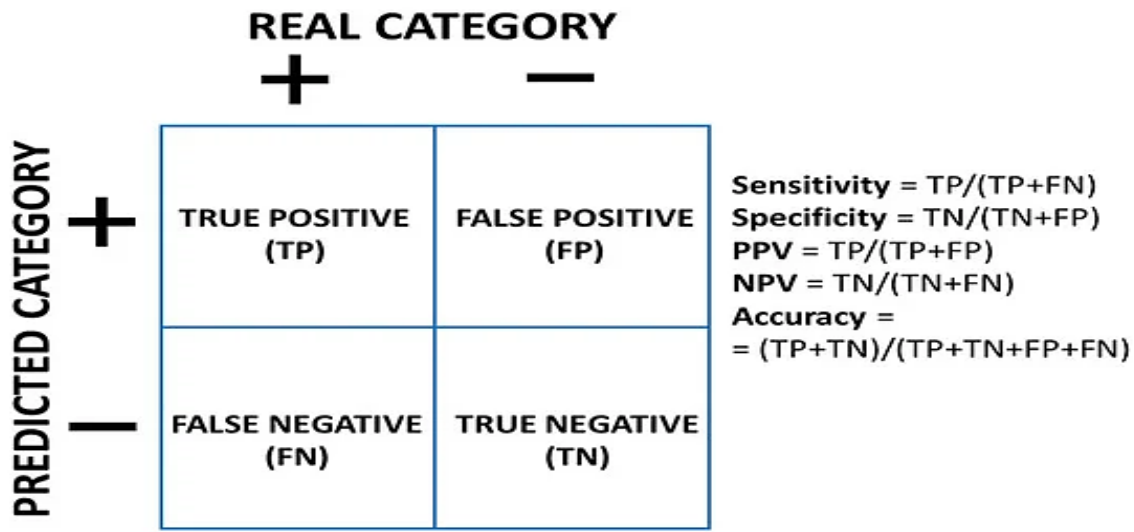

**Fig 1.** Predictive/real category of "+" or "-" and its alternating category of "+" or "-", which showcases data sensitivity, specificity, PPV, NPV, and data accuracy. Fig 1. also displays proportional sensitivity, predicting positive dataset while specificity reflects proportional negative data predictions. The positive predictive value (PPV) compares positive & negative predictive values to attain potential amounts of accuracy in the grouped dataset [53].

Let's assume that the accurate model is equal to 1, to subdue errors; the threshold (t) can be used to optimize data models to identify occurring & reoccurring numbers of people likely to face obesity in the future. The future prediction may help clinicians determine necessary weight loss treatment, working remedies, and when to administer alternative treatment. If we decide to lower thresholds to determine the current number of obese patients, it may increase sensitivity, increase false-positive ratios while decreasing specificity and PPV. The consequence of using the model (t) to identify false positive may reduce costs of treating non-obese patients, minimize the future rise of overweight populations, and makes the administered therapy to control suspected cases of obese people. To characterize the number of people predicted to have obesity in the future, the large (t) attributes optimized with PPV will unravel the discriminative capacity model using receiver operating characteristics (ROC) curves show below; to pinpoint sensitivity against 1-specificity and threshold values.

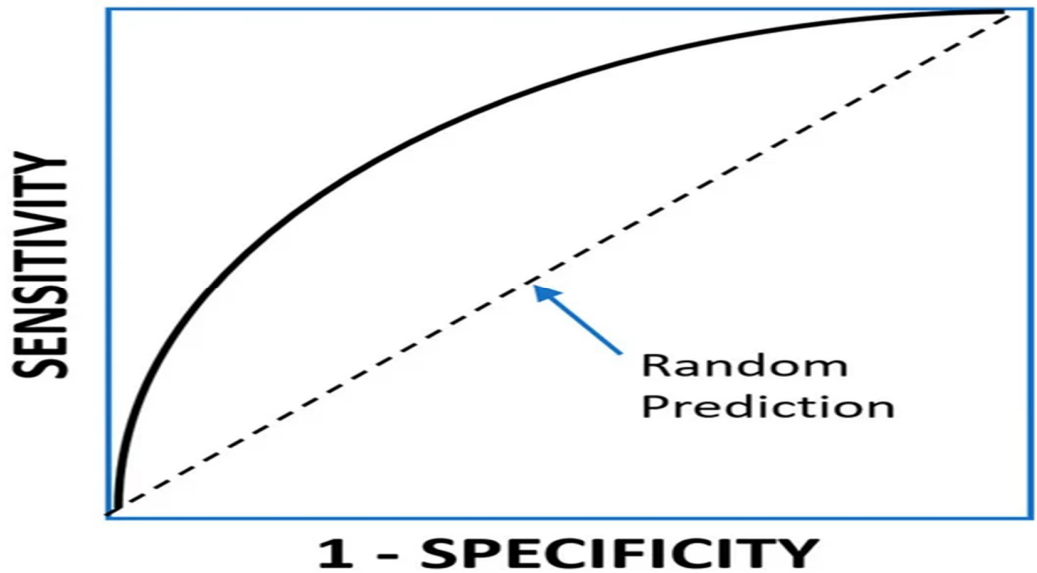

**Fig. 2** Random classified curves with diagonal numbers from (0,0) to (1,1) points have intermediating classifiers working between two extreme curve points.

The discriminatory point around the under area curve is referred to as "Area under curves of ROC Curve" (AUCROC). AUCROC perfects the adjusting real-life classification using value closer to 1 and equal to concordance value referred to as "C-Index", that predicts multi-class label datasets of two or more classes. This multi-class label predicts data performance accurately and categorizes cross-entropy (i) and m category label as equated below.

$$\sum_{J=1}^{M} lij \ Log \ pij$$

Lij- indicator variable 0 that predicts class j and showcase accurately binary number 1.

Pij- probability class predictor for j and i

N- categorical cross-entropy sum of cross-entropies numbers which measures matching probabilities of diverse classes of data frequencies. Possibly, the correlation between the predicted and actual labels using cross-entropy with errors or loss functions can be simplified with M=2 to maximize accuracy. When regression model is asserted to implement same data statistics; errors or losses may occur during training & measurement. Such error can be called "mean squared error" calculated with

$$\text{MSE} \frac{1}{n} \sum_{i=1}^{n} (yi - yi)^2$$

N- number of instance prediction

Y-yi actual continuous label

I and Yi- predictive value.

Another method of calculating error is known as "mean absolute error (MAE) equated as

$$\text{MAE} = \frac{1}{n} \sum_{i=1}^{n} (yi - yi)^1$$

Meanwhile, during data training, the model may discover too many data variables that deteriorate its performance causing model overfitting. To avoid overfitting, its significant to assess prediction performance by validating each new data inserted into existing models. To validate models, it's possible to consider internal & external validation techniques, whereas the internal validation resample and evaluate before inclusion. To achieve that, it will operate on random to resample, model fit, and evaluate data using a cross-validated bootstrap approach. Experts called the approach a "K-fold" method used to divide the total sample of subset data in random consistent folds as K and normal platforms of 5-10 folds. Each fold can be validated on an average model fitted K-1 for each K-sub-sample with each possessing k-1 folds and corresponding hold-out folds. The naïve Bayes (NB) model can alternatively serve the purpose of predicting obesity. NB rules work with approximated conditions of independent variables whilst predict variables given for each responding class. BN works under the stretch rule of probability while targeting variable y & value j to predict the conditioned variable Xi........, Xn as shown below;

$$P(Y=j/Xi, \ldots \ldots Xn) = \frac{P\left(xi, \ldots \ldots \frac{Xn}{y}=5\right)}{P(Xi, \ldots \ldots Xn)(y=5)p(=j)}$$

P(y-j)- prior probability of X from value j and p (xi............ Xn/y=j) represents the posterior probability of the conditioned variable y value j and P (xi...........xn) of the predicting variables.

The mentioned conditional probability estimate respectively, the frequencies of variable categorized and simplified with the equation below;

$$P(y=j/xi,................xn)= \frac{P\ (y=j)\pi n\ P(\frac{xi}{y}=j)}{P(xi,............xn)}$$

The class prediction set Xi.............................xn will act as a predicting value that maximize other subordinating constant factors.

K-nearest neighbors (KNN also has Xi.......Xn predictive variables assigned with constant k to training data variables less distance or similar to other datasets. KNN's major voting class assignment supports the continuous regression and value prediction to weight averagely the KNN labels. Different metrics are used to predict values and the most possible method is known as "Euclidean One", where the value of K can be variable, not constant, and heavily depends on cross-validation techniques.

Decision tree DT makes use of regression and classification method to general rectangular partitioning or space predictive variables required to split data; especially binary data and optimize its loss functions. DT labels are also assigned to partition each other to find where each label belongs to. The image below (fig 3) demonstrates how DT partition two-variable X1 and X2 uses R1, R2, R3, and R4 while generating splits S1, S2, and S3. In some scenarios, DT can be seen as a complex task because of its combinatorial binary splits and variable sequences (Breiman et al.1984)[1].

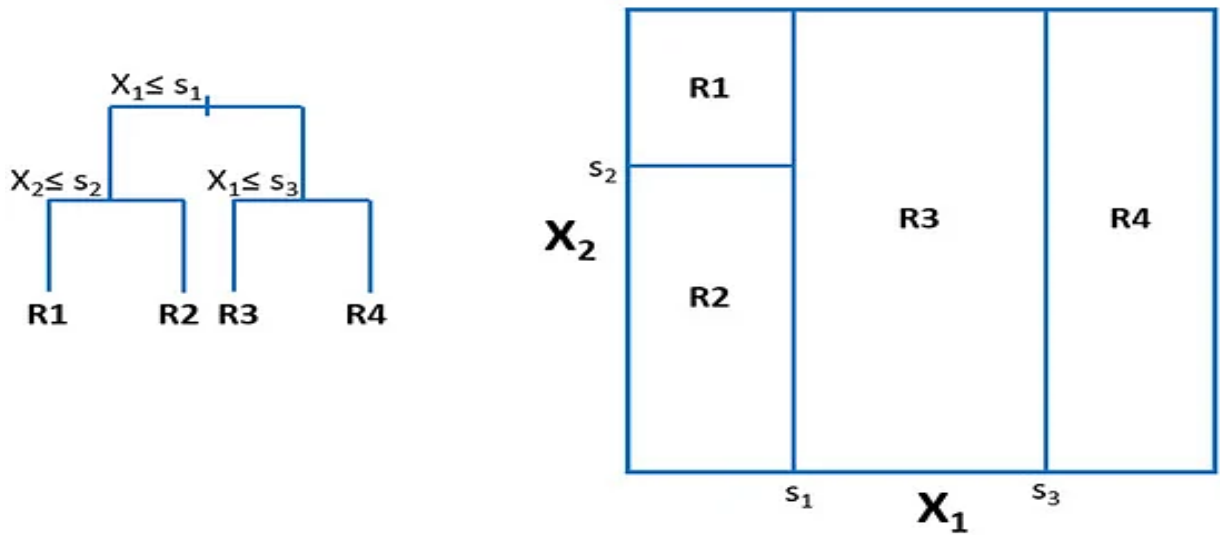

**Fig 3.** How DT partition two-variable X1 and X2 uses R1, R2, R3, and R4 while generating splits S1, S2, and S3 [53].

Quinlan (1986)[2] used CART, ID3, and C4.5 to conduct statistical tests and non-binary splits of patients with obesity. 28 developed CHIAD algorithms to interpret DT with ease and include graph aided method to display graphical high variance prediction. SVM commonly known for its binary classification can build hyperplane maximal margins to help half prediction of space result "+" and other half result prediction of "-" labels.

Maximal margins have the largest hyperplanes distance for training dataset margins. Random forest follows an ensembling method to build a high-quality aggregator mode of multiple datasets of low quality. RF ensemble controls dozens of decision tree data with little or no errors. RF is built from bootstrap called "Bagging", used to split each tree to become random subset variables and directly estimate external data validation with errors. RF train sets of DTs using B/3 approximation, where B represents the number of trees. The difference in the prediction labels is estimated with Out Of Bag (OOB) procedure.

Hence, RF suffers limitations while interpreting diverse types of DT variables, but the use of variable importance techniques will help to reduce errors and sum-up large numbers of trees. The difference between RF and gradient boost machine depends on the boost dataset, instead of using bagging methods. XG Boost Gradient ML can operate numeric data tabulation needed to implement feature data of patients with obesity. It manipulates direct complex data integrated with computer vision, speech recognition, and NLP algorithms. The regularized linear model (LASSO) is a linear residual predictor that compiles number m and predictor P. It estimated the least-square variance with test performance N<P, especially when the variance is infinite. Lasso needs a ridge regression approach to maintain betas and restrain the sum of the square betas that are less or equal to small values. BN trains obesity data with acyclic graph nodes to predict variables with more than one node or label.

**Biomarkers as a Bioinformatic approach for predicting obesity**

Obesity happens as a result of accumulated fats, excess abnormal weight gain, and compromising health risk. The deposition of disproportional fats from adipose tissue causes low-grade inflammation, adipokine deregulatory secretion, oxidative stress, altered gut hormones, and food peptides. The mentioned dysfunctional effects can cause chronic diseases like diabetes, depression, respiratory, musculoskeletal disorders (Lavie et al.2014)[3].

Scientists studied BMI to uncover measures necessary for adjusting weight, height, and body fats. BMI relates bodily fats, even underwater weighing, and dual-energy x-ray measurements are performed to determine BMI rates. BMI measurement shows beyond body fats, but discloses thick skin folds, bioelectrical impendence, and X-ray dual-energy. It shows standard waist circumference used by exercise therapists to create anti-obesity therapies and approved bariatric surgeries. (Center For Disease Center (2016)[4].

World heart federation, 2016 [5] asserted that obesity can lead to cardiovascular disease, whereas 41% of the decrease in heart-related diseases; acknowledged by therapists happens from obesity complications. Therapists used an anti-obesity scheme to promote healthier living and dissuade the habit of smoking, inactiveness, alcohol, and unhealthy diets.

WHO Globalization project (2002)[7] projects that obesity brings about bloodily inflammation, oxidative stress, and gut microbiomes. Biomarker group in 2001 recommends medical subjected heading (MESH) to evaluate normal biological, pathogenic processes or pharmacological responses to mitigate obesity.

Interleukin-6 (Ill-6) biomarker classifies obesity inflammation and relates its surrogate endpoints. It shows that obesity comes with epidemiological, therapeutic, and pathophysiology scientific evidence reflecting biochemical & biological attributes like metabolites, adipokines, and other hormones inclining obesity.

Trynka et al. 2013 [8] believe that biomarker tools can provide genetic modification and pre-disposition of obesity since genetic investigation & regulatory expression mechanisms can prevent or offer early diagnosis and protect patients from other related causes of obesity. Manna et al. 2015[9] suggest that gene-environmental interaction (GEI) provides an epidemiological study of genetic factors, diet, and lifestyle resulting in obesity. This study aids the adjustment of adipocyte growth and apoptosis that increases fat in the body. Qi et al. 2008 [10] ascertain that low pre-adipocytes reduce fat, but hypertrophy/hyperplasia increases fat causing the release of inflammatory cytokines. Trevaskis et al. 2005 [11] record different genes associated with fat accumulation such as Melanocortin-4 Receptor (MC4R and 60% obesity phenotypes, BMI, and skin-folds are considered to be inherited. Biomarkers indicate leptin levels and Loci necessary for identifying genetic markers of genomes and other related diseases.

Radiologist utilizes quantitative trait loci to detect 61 genomes through scan while observing adipogenesis and lipid turnovers causing obesity. The biomarker scan also shows insulin signals, mitochondrion, and adipokine secretion. Lohmueller et al. 2003[12] recommends prohormone convertase (1/3) PCSK and PPARG as unique genetic biomarkers for examining genetic obesity.

Genetic obesity can be associated with single nucleotide polymorphisms causing biological monogenic obesity & pathophysiological body changes. Ml biomarker known as the statistical epitasis network (SEN) provides reference computable pair-wise collaboration between genetic variants. SEN interacts with the BMI of patients to reveal twelve genes causing obesity (E.g. LEP/LEPR, GHRL, etc. (Hinuy, et al.2010) [13].

The Epigenetics heritable concordance changes human DNA sequences, influenced by environmental factors. Such DNA includes methylation of hyperguanine followed with cytokines, histone, and RNA non-coding characteristics. DNA methylation works with 55 genetic Loci known as blood leukocytes to categorize epigenetic modification and predict physical changes (obesity) and insulin resistance in the body.

DNA methylation can also influence cancer and heart disease risk. The epigenetic variance may come from environmental factors and co-play with internal genetic factors. Lovren et al. 2015[14] methylated the CPG variable as direct development or growth of obesity. Patients with abnormal DNA methylation such as Genome hypomethylation and hypermethylation are prone to chronic diseases. However, the Bioinformatics analysis of CPG islands increases obesity through CPGs density with symptoms evident in the adipogenesis of human peroxisome and other estrogen receptors.

Dick et al. 2014 [15] mentioned three CPG sites involved in the BMI analysis (CPG1, CPG 7, and CPG 5) which regulates H1F3A transcriptional adipocytes activities. Campion et al. 2010 state that an increase in H1F3A methylated levels may lead to BMI increase. Research insinuates that environmental stimulus diet and exercise can regulate inflammatory mechanisms and other contributing factors increasing H1F3A and BMI.

Stel et al. 2015[17] viewed birth obesity as an epigenomics situation while determining its mesenchyme stem cells and differentiate its Osteocytes And Adipocytes Causing fat accumulation. Research connects Micro-RNA to obesity and insists that early diagnosis & prevention can stop further complications. MiRNAs are small non-coded molecules with 21-23 nucleotide RNAs, negatively pairing genes with untranslated regions (Mansoori et al. 2015)[18]. MiRNAs contribute to body proliferation of apoptosis and regulates dozens of other genes in the body. MiRNAs are located in human tissue, serum, plasma, and blood fluids which protect endogenous RNase activities. Therefore, MiRNAs are responsible for circulating bodily fluids and its modification can correct morphological, and physiological phenotypes.

MiRans consist of five types, necessary for risk classification in patients with obesity. Ortega et al. 2010 (19) unravel MiRNAs- adipocyte mechanism to show isolated Exosomes And Microvesicles causing cancer, heart failure, and other diseases. Xie et al. 2009 [20] mentioned MIRANs level controlling Adipokines and Cytokines, which decrease obesity (MIR-103-143, and 221). Clinicians insist that MiRANs like MIR "126, 132,146, 155, and 227" can subside bodily inflammation. Clinicians use RNA non-coded numbers to mark prognosis procedures and treats bodily inflammation.non-coding RNAs and exogenous MiRNAs help the treatment of obesity by offering post-transcriptional cell that regulates & reduce stress, the proliferation of adipocyte cells, and angiogenesis differentiation. Santilli et al. 2015 added that obesity comes with excessive fatty inflammation-causing pathogenesis of insulin resistance and metabolic syndrome. The pro-inflamed cytokines lead to insulin resistance in skeletal muscles, liver, and adipose tissue. Adipokine release Resistin, Plasminogen's activator & protein inhibitor increasing body fats. However, interleukin-6 can mediate the acute protein's response by upgrading the body with CRP, fibrinogen, and Serum amyloid. (Berthier et al. 2003)[23].

Ouchi et al. 2003 [24] mentioned C-reactive protein in biomarkers for detecting plasma causing obesity and clinical epidemiological correlations between cytokines and its inflammatory factors leading to alpha tumors. Hotamisligil et al. 1994 [25] denote that tumor necrosis alpha factors (TNF) are biomarkers for controlling inflammatory response and adipose tissue secretion in the body. The activation of TNF caspases nuclear KAPPA-light (NFKB)chain enhancer can control the development of obesity since NFKB activates B cells increasing mitogen-activated protein Kinases that resist adipocytes. Experts suggest the use of glucose to counter cytokines expression in human muscles. NFKB does enhance immune response, dysregulate inflammation by forming auto-immune response preventing diseases. The use of NFKB activation can alleviate protein overexpression in B-cells, insulin, and glucose transporters (Glut & pancreas).

Harrell et al. 2015 [26] attempt to connect oxidative stress to obesity by revealing the reactive oxygen specie signals transcription the epigenetic deregulations. Stress can induce chronic low-grade inflammation, platelet activation, and endometrial dysfunction. Stress creates an imbalance between ROS (reactive oxygen species) and biological systems to detoxify repair or worn-out tissue that impacts protein lipids and DNA creating hypertrophied adipocytes.

Health & disability reports (2016)[27] mentioned lipoprotein among cholesterol factors (modified steroid) that causes obesity. Triglyceride complexity also influences human weight while increasing blood pressure, which stresses blood vessel walls, develops arteriosclerosis, and increases myocardial infarction (Stroke). Leptin belongs to the peptide adipokine's family with 6 KDa molecular weight and 146 amino acids. Its causative gene is known as OB with receptor OB-R. Leptin secretes adipose tissue causing obesity. Adiponectin secrete protein circulating in high concentration in the body. Adiponectin causes insulin sensitivity 7 lipid metabolic syndromes (Canpolat et al. 2001)[28].

Resistin contains adipokine secretion of white tissues proposed by biomarkers to cause atherosclerosis complications while Visfatin Adipokine leads to obesity, renal failure, and heart failure. Experts suggest the use of B-type Natriuretic peptide as a diagnostic biomarker to lower lipoprotein, KAPPA-B, and Lectin Density.

**Measurement & classification of obesity with ML tools**
Obesity can be measured using underweight, dual-energy x-ray absorptiometry (DEXA). DEXA measures the amount of water in the body. It can also estimate bodily electrical conductivity, body potassium, and body density. It can detect air displacement (Plethysmograph), and bioelectric impedance. DEXA was recommended for estimating bone minerals, and fat soft tissues. However, anthropometric received clinical approval as the perfect evaluating procedure for discovering the amount of body fitness, fat percentage, BMI, WC (waist circumference), HIP to waist ratio. (WHO, Global health 2009)[29]. Cadaver analysis can be used to measure body composition to discover and differentiate between essential fats & storage fats.

Wells et al. 2009[30] estimated that BMI measurement starts from 25-29.9kg/m2, where an average BMI starts from 18.5-24.9kg/m2. BMI helps to grade obesity from grades 1-3, whereas the graded number starts from 30-34.9kg/m2, grade 2 (35-39.9kg/m), and grade 3 (40 and above). Waist circumference helps to measure adipose fat amounts in the body by measuring from midpoints, minimal circumference, iliac crest, umbilicus, lowest rib, and the larger part of the waist.

**Literature review**

Falconer (1960) [61] introduced twin study techniques for estimating phenotypes heritability of monozygotic (MZ) and dizygotic (DZ) pairs of phenotypes. Both phenotypes are referred to as "Twin Repositories" sharing the same genomic square and handles the detection of obesity. Goodrich et al. (2016)[62] identified heritable Taxa gut-microbiomes' bacteria constituting to obese fatty growth. The Taxa heritable microbes influence adiposity.

Scientific inked obesity with brain addiction known as neural circuits, hereby leading to brain reward system through appetizing food & lifestyle. Obese individuals exhibit atrophy in the frontal lobes hippocampus, thalamus, and anterior cingulate cortex; more than people with normal weight. (Garcia-Garcia et al. 2015)[31].

Carnell. et al. 2012[32] predicts various impaired brain structures caused by obesity, causing obese people to seek reward through food, etc., due to a lack of cognitive control. Garcia-Garcia et al. 2013 [33] used non-invasive diffusion tensor imaging to find the correlation between the functional neuroimaging indices of obese patients. The functional indices contain amplitude low-frequency fluctuation (AlF) and regional homogeneity (Reho). The Alf controls physiological reflex and spontaneous neural activities while the Reho focuses on blood oxygen to display the level of signal fluctuation near the neuronal area.

Logan et al. 1997[34] used an executive functional stop-signal task tool to measure inhibiting control potent behavior by subtracting stop signal delays (SSD) from the time interval needed to stop fat-inflamed response. D' Elia et al. 1996 [35] measured mental flexibility and clinical decision making with MRI images to see mental reactive diffusions scan of patients with obesity.

**ML and Bioinformatics technologies**

Khader et al. 2017 [37] mentioned some scalable technologies for configured ML & Bioinformatics methods such as Apache Storm & Park, Hadoop, Mango 43.b, neo4j, and elastic for computing, integrating, and analyzing Bioinformatics data. Some data visualization tools such as Gephin, and Kibana provides visual representations of patient cases, diagnosis, and treatment. Bioinformatics ML tools such as Apache Mahout, Spark MLLib, and Weka supports ML distributed data streaming & process EMRs software for taking records of obese patients.

Chute et al. 2013 [36] unveil EMR patterns for predicting pharmacogenomics treatment of obesity. The EMR model handles data with latent variables while predicting values with latent Dirichlet allocation, model mixture, and Gaussian techniques. The techniques define low-dimensional representations of data, speed, and train predictive hidden Markov models or autoregressive dynamic time warping model. The essence is to permit new data streaming, incremental learning, and estimation of different data for different algorithms.

Ahmed et al. 2014 [38] find NN, BN, and SVM relevant for conducting data quality, false-positive ratings, and predicting health system records. Bioinformatics tools identify missing data variables that cause idiopathic conditions that influence test results. If such missing data variables are ignored, it can influence patients' quality of life, mortality & morbidity. The idiopathic condition of obese patients creates comparative diseases that transform minor ailments to complex surgical complications. Obesity according to Zipes et al. 1998 [39], leads to complex asymptotic disease centers on manifesting cardiovascular disease, stroke, heart attack, and arterial diseases. Experts captured comparable evidence relating to patients' daily activities using a time-point measurement instrument to enhance clinical decisions. The instrument controls vital signs and checks considerable symptoms like chest pain, asthma, stress test, and provoking drugs causing acute obesity & health attack.

ML & Bioinformatics commerce its biological experiments and translational therapy to support patients and intervene in the case of overweight individuals. Genome

Bioinformatics was first clinically interpreted to evaluate cardiovascular disease in obese patients Marx, 2013. The compendium variant model checks patients' genomics, proteomics, metabolomics, and transcriptomics within 14days to review a possible risk of heart failure, osteoarthritis, myocardial infarction, gait imbalance due to obesity (Clifton et al. 2014)[41].

**Enabled ML methods for predicting obesity**

An Apache Spark distribution database was integrated to coordinate subset chromosome prediction of minimum redundancy & maximum relevance (MRMR nucleotide of patients with obesity. Feature selection (FS) is an automatic sub-tagged model acting as a filter for analyzing data & decreasing memory storage requirements. FS method helps to identify or remove unwanted data features. Guyon and Clisself define "FS" as an essential tool for handling overfitting, improving model performance, and interpreting data subsets. Feature selection in ML helps biomarkers to target & differentiate between diseases & control samples. Haury et al.2011 [42] used feature selection to investigate genes selected as datasets to signature biological processes & relationships.

Steur et al. 2011 (43) used Stepwise logistic regression to validate bootstrap linear statistic models for predicting obesity. Druet et al.2012 [44] adopted a meta-analysis for predicting childhood obesity while Mayr et al (2012)[57] diagnose child obesity with quantile regression boost simulation. Nau et al. (2015)[45] trained a set size of 99 communities in the US to discover Obesogen environmental statistics using random forest.

Hasan et al. (2018)[46] weighed community obesity reduction & sedentary behavior reduction with deep learning algorithms. Oksuz et al. 2018[47] weight clinical decrease in the use of therapy for obesity with KNN, SVM, and GBM. Tueir et al (48) 2018 reviewed physician diagnostic successful treatment of overweight children with the Adhoc algorithm whilst Duran et al. (2019)[49] measured excess body fat percentage with the ANN tool. Guruswami and Sahai, 1999 [50] utilized adaptive boosting (AdaBoost) ML algorithm to solve multi-class problems, train weak datasets, control overfitting, and poor performance.

The University of Michigan nutrition obesity research center conducted a weight management program to device means of adding multidisciplinary lifestyle intervention to promote 15% weight loss within a few weeks.

Genome-Wide Association Studies (GWAS) revealed genomic Loci variants with complex traits to evaluate the number of people with obesity with the aid of feature selection. GWAS database generated results are compared with ML algorithms to show BMI and waist circumference and categorize genomic patients underweight, normal weight, and overweight. To solve data imbalance problems, normal & risk classification methods are used to check obese, and extreme obese categories. The genetic profiles of patient datasets show patients age, gender, height, weight; taken into consideration during data prediction. A machine learning approach utilized for training & testing the datasets can also be automated to perform hyperparameter optimal configurations for each classified data. The tables below shows dataset variables, data description, RF algorithm with feature selected model; classifying disease traits, and obesity
Detection.

**Table 1:** Variables/Description of Status, Feature Selection (FS) analysis RF [64]

| Variables | Description |
|---|---|
| Age | Age |
| Gender | Male = 0, Female = 1 |
| Height | Height in meters |
| Weight | Weight in Kg |
| BMI | $\frac{\text{Weight (Kg)}}{(\text{Height(m)})^2}$ |
| Status | Underweight, normal range, Overweight, Obese and Extremely obese |

LIST OF FEATURES SELECTED USING THE RF ALGORITHM

| Feature | Chrom | Reported gene/s | Strongest.SN P.risk.allele | Disease-trait |
|---|---|---|---|---|
| rs12567355 | 1 | CD53 | rs12567355-A | Obesity-related traits |
| rs3001167 | 1 | EEF1A1P14 | rs3001167-G | Obesity-related traits |
| rs586688 | 1 | NAV1 | rs586688-A | Obesity-related traits |
| rs7525133 | 1 | RHBG | rs7525133-A | Visceral adipose tissue adjusted for BMI |
| rs17104630 | 6 | NKX2-1 | rs17104630-G | Height |
| rs12970134 | 10 | MC4R | rs12970134-A | Type 2 diabetes, BMI, Weight, Waist circumference |
| rs12978500 | 11 | C2CD4C | rs12978500-C | Obesity-related traits |
| rs10195252 | 12 | COBLL1, GRB14 | rs10195252-C and -T | Triglycerides and Waist-hip ratio |
| rs11241130 | 18 | NREP | rs11241130-G | Obesity-related traits |
| rs2076529 | 19 | BTNL2 | rs2076529-C | Waist-hip ratio |
| rs1447295 | 21 | MYC, intergenic | rs1447295-A | Prostate cancer |
| rs4242382 | 21 | intergenic | rs4242382-A | Prostate cancer |
| rs10090154 | 21 | NR | rs10090154-T | Prostate cancer |

**Table 2:** Parameters, Classifiers, Categories & object class [64]

TUNING PARAMETERS SELECTED

| Object class | Parameters | Best |
|---|---|---|
| gbm | n.trees interaction.depth shrinkage n.minobsinnode | n.trees = 100 interaction.dedepth = 1 shrinkage = 0.1 n.minobsinnode = 10 |
| glmnet | alpha lambda | α = 0.1 λ = 0.02768717 |
| rpart | cp | cp = 0 |
| knn | k | k = 7 |
| svmRadial | C sigma | C = 0.5 σ = 0.04797839 |
| rf | mtry | mtry = 2 |
| nnet | size decay | size = 3 decay = 0.1 |

OBJECT CLASSES SELECTED

| Object class | Classifier | Category |
|---|---|---|
| gbm | Stochastic Gradient Boosting. | Nonlinear |
| glmnet | Lasso and Elastic-net regularized generalized linear models. | Linear |
| rpart | CART (Classification and Regression Trees). | Nonlinear |
| knn | K-Nearest Neighbor. | Nonlinear |
| svmRadial | Radial Basis Function Kernel Support Vector Machine. | Nonlinear |
| rf | Random Forest. | Nonlinear |
| nnet | Backpropagation Neural Network. | Nonlinear |

**Recommendations**

Principles of clinical treatment believe the precision medicine is essential for handling bio-medical investigations of genomics and physiological disease progression. Researchers studied electronic medical records (EMRS) containing medications and precision therapy records of patients recovering from obesity. Bioinformatics involves the comprehension of molecular, social, physiological, and environmental states of health leading to chronic ailments.

The Bioinformatic study deals with the evaluation of disease phenotypes necessary for alleviating comorbidities, making it essential to use real-time data streams to capture, analyze, aggregate, and visualize data types, source, and integrating devices for translational activity in Bioinformatics. Bioinformatics data in health care known as Compendium data aggregate can monitor generated data used in EMR software to check ambulatory, prognosis, intervention, and stratification of patients diagnosed with obesity.

The hospital ensures to generate fitness records using wearables and biosensors to track treatment & recovery. The biosensors authenticate patients' heart rate, glucose variability while disclosing respiratory, echocardiography, biological, and other vital signs. The profiled biological experiment crosschecks genome and exome sequence metabolic biomarkers of obese patients. Apple Research Kit helps healthcare to integrate data using an application like iPhone to track walking, and Photoplethysmographic heart rates of obese patients. The tool can offer temporal monitoring of fluctuating heart rate of patients. (Berwick, 2008) (51). Biobank acts as a bio-respiratory & biological store for molecular sampling, collection of materials, process, and bio-specimens for obese patients. Diverse Bioinformatic databases like bio-GPS and KEGG aids molecular classification of obese disease mechanisms. For instance, the use of 3D domain/Swap as core molecules for protein study & monitoring of physiological records of obese individuals. . (Goldberger et al. 2000)[52].

Lukowicz (2004) (54) considered wearables relevant for monitoring lifestyle and tracking fitness measures (e.g Steps climbing and distance covered during running & walking, gait, and posture). Wearables offer a relative analysis of heart rate, calories burn, and sleep quality; as well as calorie intake, computed score of the progress of patients.

Takacs et al. 2014 (55) insist that wearables won't offer extensive validating metrics for monitoring blood pressures, respiratory, blood oxygenation.

European society of hypertension designed similar subject devices to observe blood pressure, blood oxygenation, and respiratory level of patients (Wilhings, 2015)[56]. medical technologist developed a featured application with Bioinformatics data recording to keep track of patients recovering & treatment progress. Basis B1 waistband works as an accelerometer for controlling heart rate, ambient & body temperature, skin conductance, and caloric burn recorder for patients struggling with obesity. Body Media Link Alms controls heat flux, body motion, skin conductance while promoting sleep & calorie burn. Fitbit aria helps BMI & weight monitoring whilst Fitbit surge serves as GPS & altimeter devices for monitoring daily activities & progress of obese patients. The hexoskin smart skin gives clear records of obese patients ECG, respiratory rate & tidal volume while promoting sleep & posture. IHealth BP5 Glucometer controls blood pressure & glucose level. Jawbone-ups relate heart rate, accelerometer, caloric intake, and skin temperature. Mapmyfitness records food intake & activity monitoring. Scanadu Scout offers support for handling blood temp pressure, blood oxygenation, and ECG monitoring (Custodia, et al. 2012). (58).

**Conclusion**

In the US, Harvard estimated up to 65% increase in overweight individuals (500 million adults), caused not only by living an unhealthy lifestyle, but stress that increases blood pressure & adipocytes and contributes to endothelial dysfunction. ML applications can help to predict obesity and practically low-cost & continuous increase. Yong et al.2011 [63] indicates that BMI, Waist & hip ratio can lead to obesity & other health problems, suggesting the use of linear, logistic regression and ROC curve to analyze epidemiological variables of BMI, WC, and WHR. This report has been reviewed using different ML algorithms & techniques for predicting obesity. The research centers its emphasis on Bioinformatics tools used to investigate weight gain and problem suggestions on how to achieve weight loss. It studied genetic and medical data analysis used to attain specificity & sensitive rating of patients with obesity.